\documentclass[aps,pre,twocolumn,superscriptaddress,showpacs,floatfix]{revtex4}
\usepackage{graphicx}
\usepackage{epsfig}
\newcommand{\bs}{\symbol{92}}

\begin{document}


\title{Optical structure and function of the white filamentary hair covering the\\edelweiss bracts}

\author{Jean Pol Vigneron}
\email{jean-pol.vigneron@fundp.ac.be} \affiliation{Laboratoire de
Physique du Solide, Facult\'es Universitaires Notre-Dame de la
Paix, \\61 rue de Bruxelles, B-5000 Namur, Belgium}

\author{Marie Rassart}
\affiliation{Laboratoire de
Physique du Solide, Facult\'es Universitaires Notre-Dame de la
Paix, \\61 rue de Bruxelles, B-5000 Namur, Belgium}

\author{Zofia V\'{e}rtesy}
\affiliation{Research Institute for Technical Physics and
Materials Science, POB 49, H-1525 Budapest, Hungary}

\author{Kriszti\'an Kertesz}
\affiliation{Research Institute for Technical Physics and
Materials Science, POB 49, H-1525 Budapest, Hungary}

\author{Micha\"{e}l Sarrazin}
\affiliation{Laboratoire de
Physique du Solide, Facult\'es Universitaires Notre-Dame de la
Paix, \\61 rue de Bruxelles, B-5000 Namur, Belgium}

\author{Laszlo P. Bir\'{o}}
\affiliation{Research Institute for Technical Physics and
Materials Science, POB 49, H-1525 Budapest, Hungary}

\author{Damien Ertz}
\affiliation{National Botanic Garden of Belgium,
Domein van Bouchout, B-1860 Meise, Belgium}

\author{Virginie Lousse}
\affiliation{Laboratoire de Physique du Solide, Facult\'es
Universitaires Notre-Dame de la Paix, \\61 rue de Bruxelles,
B-5000 Namur, Belgium} \affiliation{Department of Electrical
Engineering, Stanford University, Stanford, California 94305, USA}

\date{\today}

\begin{abstract}
The optical properties of the inflorescence of the high-altitude \textit{Leontopodium nivale} subsp. \textit{alpinum} (edelweiss) is investigated, in relation with its submicrometer structure, as determined by scanning electron microscopy. The filaments forming the hair layer have been found to exhibit an internal structure which may be one of the few examples of a photonic structure found in a plant. Measurements of light transmission through a self-supported layer of hair pads taken from the bracts supports the idea that the wooly layer covering the plant absorbs near-ultraviolet radiation before it reaches the cellular tissue. Calculations based on a photonic-crystal model provides insight on the way radiation can be absorbed by the filamentary threads.
\end{abstract}

\pacs{42.70.Qs, 42.66.-p, 42.81.-i, 42.81.Qb, 87.19.-j}

\maketitle

\section{Introduction}
\begin{figure}[b]
\centerline{
\includegraphics[width=8.0cm]{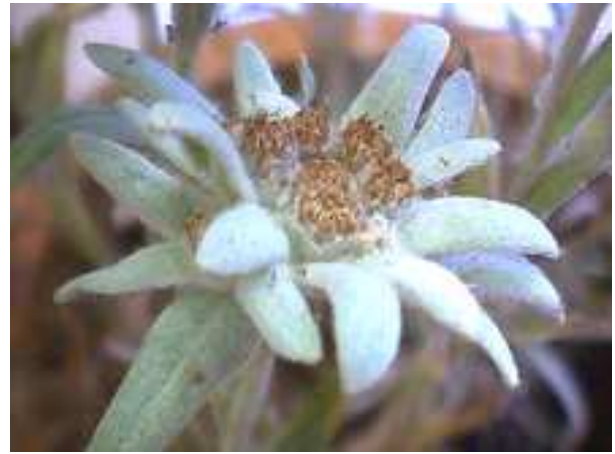}}
\caption{(Color online) \textit{Leontopodium nivale} (edelweiss) are herbaceous plants which develop a characteristic star of silver-white foliaceous bracts particularly dense on the side facing the sky. The reflectance of this wooly layer is high at all visible wavelengths but, as demonstrated in the present work, shows a strong deficiency in the near ultraviolet range. Picked up too frequently in the last centuries, the edelweiss has become a rather rare plant, with the consequence of a protection in most countries where it can be found. All material used in the present work has been taken from cultivated specimens.} \label{fig1}
\end{figure}

\textit{Leontopodium nivale} subsp. \textit{alpinum} (Cass.) Greuter ($\equiv$ \textit{Leontopodium alpinum} Cass.)\cite{Greuter-W-2003} is a perennial herbaceous plant which lives in european mountains, where it can be found (in the Alps) up to an altitude of 3400 m. The whole plant, including stems, leaves and bracts surrounding flowers are abundantly felted with white hairs. The inflorescence is composed of two to ten small capitula crowded together at the apex of the stem and subtended by a star of five to nine densely white-downy foliaceous bracts (see Fig. \ref{fig1}). The white hair covering the plant is thought to limit water evaporation because the plant distributes over very dry and windy regions. The plant is indeed particularly resistent to drought. The cells which make the living tissue of the plant are known to be an absorber of ultraviolet (UV) radiation. It was even suggested in a recent patent\cite{martin-brevet-2003} that this could be used in a preparation produced from dedifferenciated cells grown \textit{in vitro} for the protection of the human skin against agressive ultraviolet radiation. The real benefit of the compounds found in the genus \textit{Leontopodium} cells for cosmetic or medical applications\cite{dobner-je-2003} is still a matter of discussion, but the sensitivity of these cells to penetrating UV radiation raises quite interesting questions. One of these is an issue of protection\cite{Starr-book-2004} : how does the plant resist the high flux of the energetic and harmful irradiation to which it is exposed in the high-altitude rarefied atmosphere?

The way ultraviolet light is handled by the plant may be better understood if we consider in more detail the structure of the wooly layer covering the foliaceous bracts surrounding the inflorescence. However, in this case, the optical properties of this filamentary pads seem to be determined in part by its submicronic structure, so that examination of these objects using scanning electron microscopy is required. This observation provides an opportunity to demonstrate that this plant carries an optical structure which, from many points of view, is reminiscent of recently described artificial photonic-crystal optical fibers\cite{knight-OL-1996,knight-OL-1997,joannopoulos-pup-95}. The short-range order and long-range disorder found in this wooly material somewhat complicate the identification of the mechanisms which lead to the energy absorption but, as an attempt of clarification, we will report on optical reflection and transmission experiments, focusing on the properties of the thin wooly coating of the plant, rather than on the cells forming the living tissue. Simulations will be carried out in support of the understanding of the outcome of these measurements. The cultivated specimen of \textit{Leontopodium nivale} used in this study has been obtained from the \bs Jardin Alpin du Lautaret'', France.  It is kept in the herbarium of the National Botanic Garden of Belgium (BR~; Vigneron~\&~Ertz n$^{o}$7065).

\begin{figure}
\centerline{
\includegraphics[width=8cm,height=6cm]{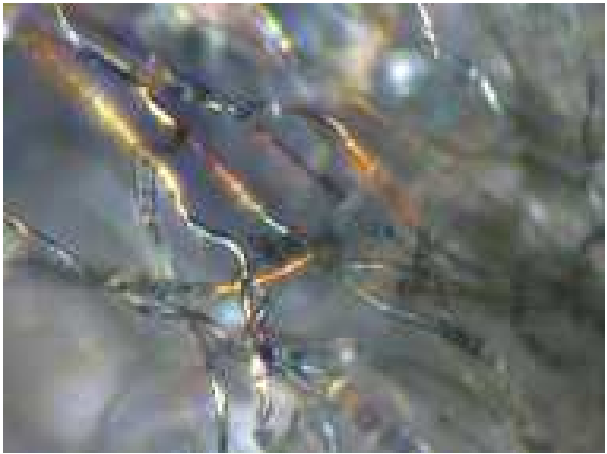}}
\caption{(Color online) Optical microscope image of the criss-cross of transparent filaments which form the white cottony pads covering the edelweiss bracts. The image was recorded in transmitted polarized light. The diameter of the filaments can be estimated near 10 $\mu$m in diameter. A slight iridescence can be noted.} \label{fig2}
\end{figure}
\begin{figure}
\centerline{
\includegraphics[width=8cm,height=6cm]{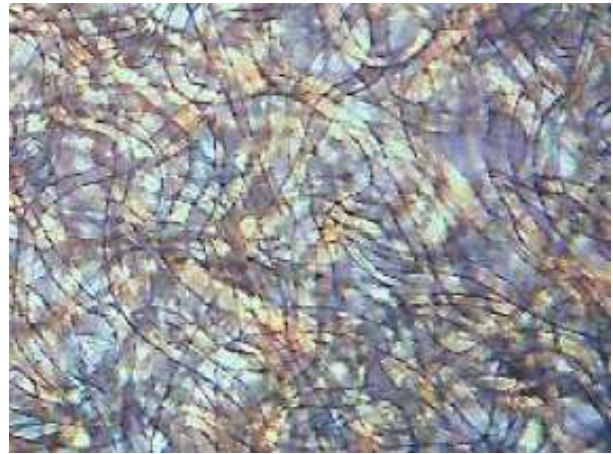}}
\caption{(Color online) Optical microscope image of the white pads, detached from the edelweiss bract, suspended in water and illuminated in transmission. The filaments clearly appear transparent. From this image, it can be inferred that the transparent materials making the filaments has a refractive index very slightly higher than that of water.} \label{fig3}
\end{figure}

\section{Microscopic structure of a \textit{Leontopodium} pubescent bract surface}

The surface of the bracts is covered by a macroscopic layer of white cottony pads. The thickness and density of this hair layer is highly variable. Our experiments were conducted on cultivated samples with a rather low hair density. The optical microscope easily reveals the disordered criss-cross structure of the transparent filaments which forms this white pads layer. The image on Fig. \ref{fig2} is taken on the border of a bract, where the filaments are rarefied, under transmission illumination. A linear polarization of the illuminating light resulted in an increase of the image contrast. This polarized image also shows some iridescence, which may indicate an optical structure at a finer scale. Fig. \ref{fig3} shows a slightly different view of the criss-cross structure. Here, the white pads has been detached from the bract, and observed as a suspension in water. The filaments are clearly seen to be transparent, with a refractive index not very different from that of water. This picture indicates that the material which makes the filament exhibits a refracting index close to 1.4, a value which later in this work, will be used to support simulations.

\begin{figure}
\centerline{
\includegraphics[width=8cm]{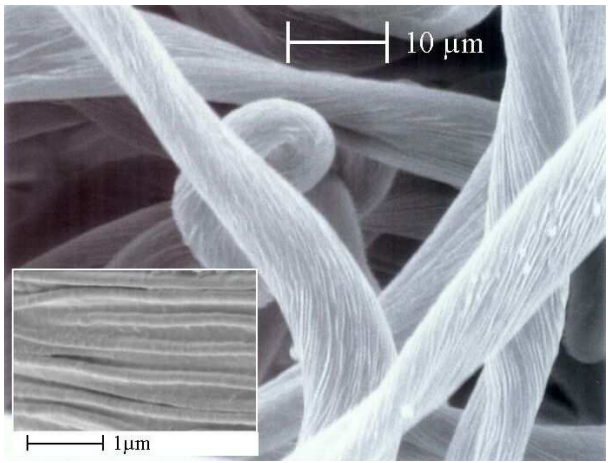}}
\caption{Scanning electron microscope picture of the filaments covering the leaflets surrounding the \textit{Leontopodium nivale} bracts. Note the submicron structure of the filaments (inset) : with a high-magnification the filament surface appears to be outfitted with an array of thin parallel fibres, about 0.18 $\mu$m in diameter.} \label{fig4}
\end{figure}

Optical properties are however not determined at this length scale. In order to better know how light waves behave in the white pads, it is necessary to obtain information at magnifications showing submicron details. A scanning electron microscope (SEM) can be used for this purpose. Fig. \ref{fig4} shows the filaments under the SEM, revealing their longitudinal structure.

\begin{figure}
\centerline{
\includegraphics[width=8cm]{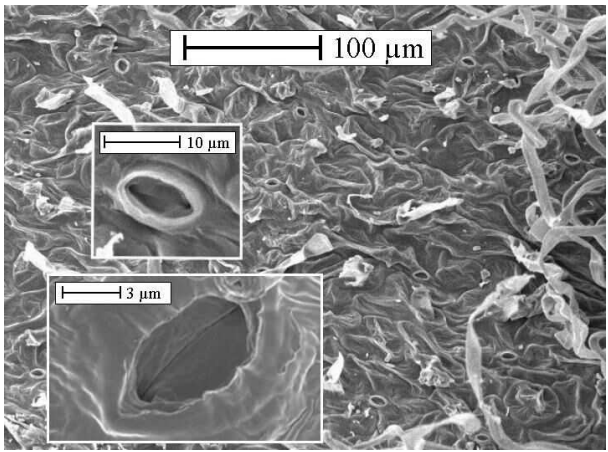}}
\caption{Scanning electron microscope picture of the remains of the filaments after the hairs have been cut out. The base of each filament appears as a small crater, indicating that the filament is a hollow structure (upper inset). The lower inset shows the bottom of the scar left by the removal of a hair.} \label{fig5}
\end{figure}

The filament transverse section does not show a constant shape: the changes of apparent diameter along the length of the filament actually correspond to rather strong deformations. The filaments appear to change their section shape along their length. This is only possible because the filaments are found to be hollow. This has been confirmed by carefully shaving the bract surface and examining the remaining stump (see Fig. \ref{fig5}). This observation indicates a wall thickness below the micrometer significantly smaller than the filament diameter.

The second observation is the transverse structuring of the filament, whose surface clearly shows a peripheral array of parallel fibres. The diameter of these fibres is of the order of 0.18 $\mu$m, the order of magnitude of the wavelengths of near ultraviolet radiation. The filament is then possibly a photonic structure whose optical properties may influence the UV reflectance and transmittance of the hair layer covering the bract.

In the longitudinal direction, the spatial structure varies slowly, when compared to the variations found in the transverse direction. A reasonable view of this structure is then to consider it as a two-dimensional photonic structure, though recognizing that, in reality, some roughness exists along the \bs invariant'' direction. A simplified view of the filaments which constitutes the white hair layer considers two distinct length scales : the large scale structure, with typical inhomogeneities spanning 10~$\mu$m or more, and the small scale structure, with typical lengths well under a micrometer, which appears as a much more ordered structure reminiscent of photonic-crystal optic fibers.

\begin{figure}
\centerline{
\includegraphics[width=8cm]{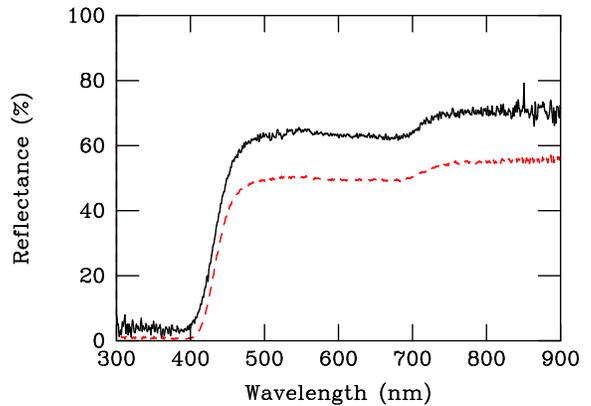}}
\caption{(Color online) Reflectance of light, for normal incidence, on an edelweiss bract. The dashed line reports a specular reflection, while the solid line reports a reflection integrated over all outgoing directions. The reflectance is nearly constant over the whole visible range of wavelengths (explaining the white appearance of the bracts). It is found very weak in the ultraviolet range, just below 400 nm. In this range, the bracts do not transmit any wavelength, so that the reflectance deficiency indicates a global radiation absorption.} \label{fig6}
\end{figure}

\section{Reflectance and absorption in a complete edelweiss bract}

The reflectance spectra of a complete leaflet taken from the star-like organ surrounding the edelweiss flowers have first been investigated using an Avaspec 2048/2 fiber optic spectrometer. Measurements were performed, under quasi-normal incidence, in the specular direction and, using an integration sphere, accumulating all scattering directions. The results are shown on Fig. \ref{fig6}. Specular and integrated reflectances are very similar and show a constant value over the whole visible range, except for a slight jump observed near 700 nm. This spectral profile explains in a very direct way bright-white appearance of the bracts. But the most spectacular reflection variation occurs when entering the near-ultraviolet region, where it is found that no light is actually back-scattered.It is important to note that bracts are rather thick leafs and, in transmission, they turn out to be opaque to all optical radiations. The missing ultraviolet radiation is then actually absorbed and this may be harmful to the plant, if the ultraviolet light penetrates into the living cells of the leaf and produces damage there.

\begin{figure}
\centerline{
\includegraphics[width=8cm]{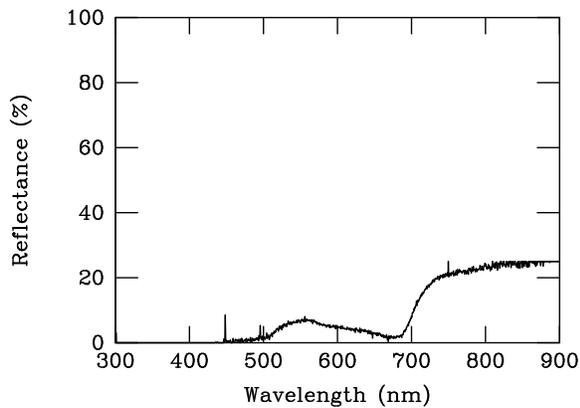}}
\caption{Light reflectance, for normal incidence, on an edelweiss bract removed from its hair layer. The spectrum, collected in an integration sphere, shows the characteristic green reflection of vegetal cells. Absorption can take place at other wavelengths, including the near ultraviolet range, below 400 nm, exposing the cells to radiation damages.} \label{fig7}
\end{figure}

\begin{figure}[b]
\centerline{
\includegraphics[width=8cm]{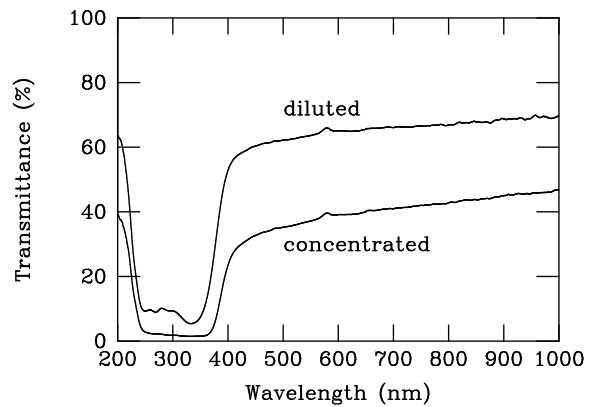}}
\caption{Transmittance of an extract from the cells found in the bracts living tissue. The \bs diluted'' preparation is a mix of the \bs concentrated '' extract with an equivalent volume of ethanol. The measurement is done through a quartz cell offering an extended transparency in the UV range. The normalizing intensity refers to pure ethanol tested in the same conditions.} \label{fig8}
\end{figure}

An important issue is then to know whether, in nature, the absorption takes place preferentially in the cells or in the wooly hair on the bracts. We have tried answering this question by separating the green living part of the bract from the white hair layer and perform separate optical measurements on these isolated materials. As shown on Fig.  \ref{fig7}, the green leaf substrate essentially reflects the green band left out with the living-vegetal chlorophyll activity. Note also the reflection band for wavelengths larger than about 700 nm, which can explain the jump of reflectance found on Fig. \ref{fig6}. The surface of the bare leaf is ready to absorb other wavelengths, including the near ultraviolet radiation.  This absorption capability could be confirmed by preparing a bracts extract, using a procedure described in Martin\textit{ et al.}\cite{martin-brevet-2003}. Fresh bracts wetted with ethanol, ground in a mortar, were first filtered using Filtrak 1388 filter paper. The transmission spectrum extract(Fig. \ref{fig8}) was recorded after a few days of chlorophyll photodegradation. The extract could be dried and re-dissolved in ethanol without altering the transmission spectrum. This measurement confirms the sensitivity of the living cells to the UVA radiations.

The observation of absorption in the leaf cells justifies the assumption of possible chemical damages occurring at these UVA wavelengths, so that the plant does need a protection in the high-altitude environment. In the next section, we investigate more specifically the possible shielding offered by the wooly hair layer covering the bract.

\section{Reflectance and transmittance of the layer of white filamentary pads}

\begin{figure}
\centerline{
\includegraphics[width=8cm]{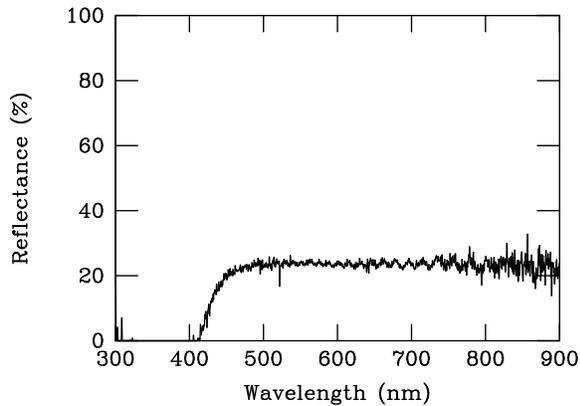}}
\caption{Reflection from an isolated layer of white pads deposited on a glass substrate. The reflectance was measured using an integration sphere at near-normal incidence. Note the reflectance deficiency in the near ultraviolet region.} \label{fig9}
\end{figure}

\begin{figure}
\centerline{
\includegraphics[width=8cm]{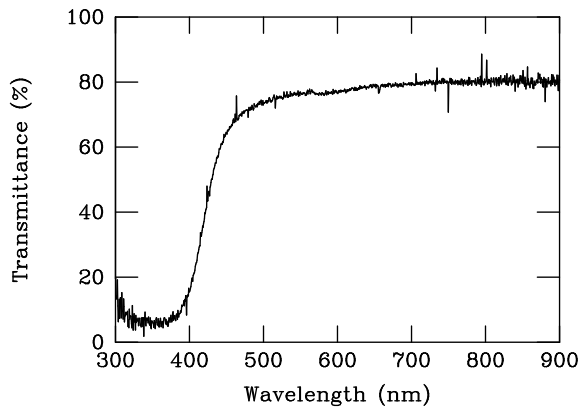}}
\caption{Transmission through an isolated layer of white pads deposited on a glass substrate. The transmittance was measured using an integration sphere at near-normal incidence. Note the transmittance deficiency in the near ultraviolet region. Deficiencies in transmittance and reflectance at the same wavelengths implies, ultimately, absorption at these wavelengths.} \label{fig10}
\end{figure}

The layer of cottony hair covering the bract was carefully removed and rearranged on a glass plate so that the thickness and the density of the layer do not exceed those found on the living plant. Reflectance and transmittance measurements were then performed, using an integration sphere. By recording the transmission of the bare glass slides under identical conditions we checked that the supporting glass did not induce detrimental absorption till 300 nm UV radiation.  The reference beam was defined in presence of the glass support, so that any variations of optical properties due to the wavelength dependence of the glass substrate was corrected for. The result is shown on Fig. \ref{fig9} and Fig. \ref{fig10} for reflectance and transmittance, respectively. The distribution of the reflectance over the near-ultraviolet and visible spectra does not differ much from that obtained \textit{in vivo}, when the filaments layer rests on the naked plant leaf. The reflectance is still constant over the whole visible spectrum, above 400 nm. The same disappearance of reflectance is also noticed in the ultraviolet band between 300 and 400 nm. The preparation could be kept for several weeks without noticeable changes of its optical properties.

Even more instructive, we find the same kind of filtering function in transmission. The transmission through the wooly layer is rather uniform over the whole visible range but presents a pronounced dip in the ultraviolet region between 300 nm and 400 nm. The deficiency of reflectance and transmittance in this range of ultraviolet radiation means that a strong absorption occurs there. Note that by \bs absorbed'' we mean radiation which looses energy while transiting through the filament wall, but also, eventually, radiation which first turns into guided waves in a Fano resonance process (see below) and then disappears with the lifetime of the guided mode. In the visible part of the spectrum, the sum of reflectance and transmittance is close to $100\%$, so that very little absorption is actually expected. The wooly layer tends to protect the plant from near ultraviolet (UVA, from 320 nm to 400 nm) but does not hinder the exposition to visible radiation needed to sustain biological processes. The attenuation of the visible spectrum is essentially controlled by the reflection and not by the absorption.

The sources and setup used in the present experiments did not provide access to other ultraviolet ranges (UVB, from 290 nm to 320 nm and UVC, from roughly 100 nm to 290 nm) so that we do not bring any conclusion for these ranges of deeper UV radiations. In order to better understand the mechanisms involved in the optical properties just described, we performed a series of numerical simulations based on a simple model of a filamentary surface. These simulations are reported in the next section.

\begin{figure}
\centerline{
\includegraphics[width=8cm]{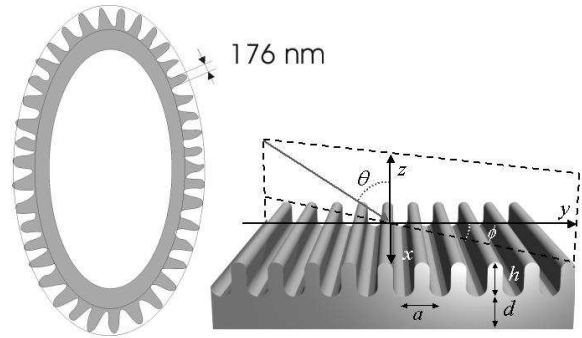}}
\caption{Filament cross section model, as deduced from field-emission scanning electron microscopy. The structure is a hollow tube, about 10 $\mu$m in diameter. An array of well-separated parallel fibers with a lateral diameter of $176$~nm is attached to the external surface of the tube. In the calculations, the curvature of the wall is neglected. The wall is viewed as a grating which is shown on the right. The grating period is taken to be $a=420$ nm. The other parameters $h=410$ nm and $d=400$ nm have less incidence on the grating optical properties. The refractive index of his structure is chosen to be 1.4. The incidence plane orientation with respect to the direction of the grating corrugation is measured by the angle $\phi$, as shown.} \label{fig11}
\end{figure}

\section{Theoretical model for the absorbance of filamentary surfaces}

The structure described by the scanning electron microscope is highly complex because it exhibits inhomogeneities at very different length scales. The submicronic fibres in the filament is only a fraction of micrometer in diameter; the filament itself has a diameter of about ten micrometers, and the filaments form a disordered medium which extends over distances a thousand times larger. We will not attempt to produce a model which accounts for all aspects of light scattering at these various length scales, but rather focus on the role of the submicronic surface corrugation on the radiation absorption.

We then only consider the surface structure of the filament, which we assume perfectly ordered. The filament is actually a tube, with a wall a fraction of a micron thick (see Fig. \ref{fig11}). The tube has a diameter of about 10 $\mu$m, so that the wall has a rather large curvature radius. The curvature can then be neglected, and we end up representing the wall as a planar slab bearing a grating-like surface corrugation. As the inset of Fig. \ref{fig4} shows, the submicronic structure of the filament surface can be described as a collection of parallel fibers with a very small and constant diameter, attached on a more or less flat surface. The examination of these fibers allows to estimate their average diameter, 0.176 $\mu$m. The refractive index of the cylinders is taken to be 1.4, slightly higher than that of water in the visible. The geometry of the model is summarized on Fig. \ref{fig11}. From the filament cross-section sketched on the left, the ideal grating profile is constructed and shown in perspective, on the right inset. The curved parts of this profile are semi-circles and the protrusions width is given the value 0.176 $\mu$m, as mentioned before. Keeping the same typical dimensions, a slightly different profile (using a gaussian shape of the protrusions) has also been studied, with the conclusion that minor modifications of the profile could not bring any significant change in the grating response.

We focus the investigations on the reflection of visible and ultraviolet electromagnetic waves. It should be repeated that this model does not contain all features shown on Fig. \ref{fig4} and that characteristics such as the fibre longitudinal irregularities or the randomness of the filament orientation are not accounted for.

The reflectance of this photonic structure is computed using a transfer matrix method in a plane wave representation. A brief account of the method has been presented elsewhere\cite{Lousse-PRB-2004,pendry-PRL-1992,lousse-thesis-2003} so that we will not recall any technicality about these computations.

\begin{figure}
\centerline{
\includegraphics[width=7.7cm]{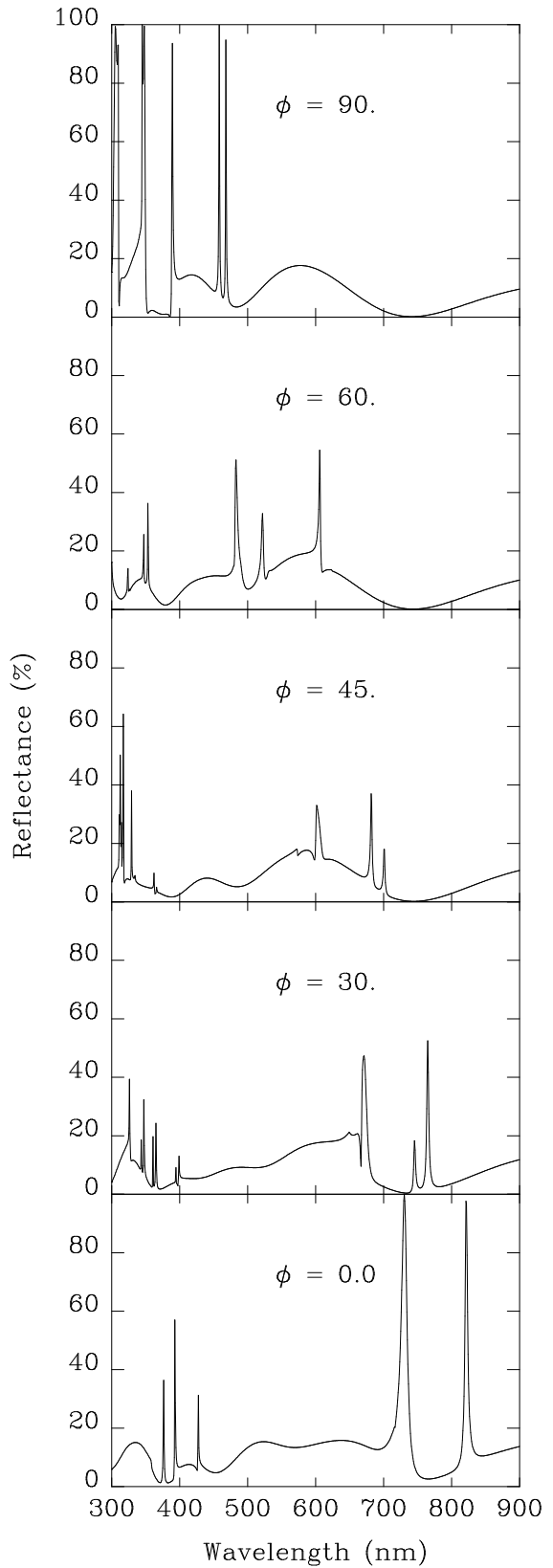}}
\caption{Reflectance of a slab structured as shown on Fig. \ref{fig11}, for a light wave impinging at 45$^{o}$ from normal in an incidence plane perpendicular which makes various angles $\phi$ with direction of the grating corrugation (TE polarization).} \label{fig12}
\end{figure}

We consider a monochromatic plane wave impinging on the wall surface. The incidence plane crosses the fibres at right angle and the angle of incidence (from the normal) is a typical 45$^{o}$. In all the results described here, the transverse-electric (TE) incident-wave polarization is presented. Transverse-magnetic (TM) polarization was also examined but will not be reported in detail : the spectra are very similar and this analysis brings no further insight. The reflectance is shown on Fig. \ref{fig12}, as a function of the incident wavelength for several orientations of the incidence plane ($\phi$ is the angle made by the incidence plane with the direction of the corrugation). These spectra show rather sharp structures superimposed to a broader background which reaches a high value of about 20\%.

For our purpose, the most noticeable part of the spectrum is the near-ultraviolet region, between 300 and 400 nm. The sharp, asymmetric lines which appear there are related to the Bragg scattering of modes which conduct light through the network of peripheral fibres. Long-lived guided modes are normally not seen in the reflectance spectrum of a flat film, as they are associated with evanescent waves in the incidence and emergence regions surrounding the film. As is well known the separation of guided modes and radiative modes in a planar slab or a perfect wave guide is a consequence of the full translational invariance of the film when displaced parallel to itself. A lateral roughness makes this separation inappropriate, and allows some degree of hybridization between these two distinct classes of modes. As a consequence, the sharp guided modes become leaky and, conversely, can be loaded with energy by an external radiation. It was shown\cite{fan-PRB-2002} that the superposition of the guided photonic modes with the non-resonant transiting radiation give rise, in the transmission and in the reflection spectra, to asymmetric lines whose shape has been  referred to as Fano profiles\cite{fano-PR-1961}. The presence of these easily recognizable structure denote the possible energy transfer from incident waves into guided modes.

As seen on Fig. \ref{fig12} these modes accumulate in the short-wavelength, near-UV, part of the spectrum, disturbing the regular pattern of Fabry-Perot resonances. The presence of these peaks in the UV part of the spectrum simply means that the \bs 176 nm'' fibers attached to the external surface of the filaments supports UVA guided modes and that, because of the periodic lateral corrugation, they can be excited by the incident wave. As seen on Fig. \ref{fig12}, the presence of these modes in this narrow spectral region is maintained there whichever the orientation of the incidence plane. It can then be expected that the process of energy transfer into UVA guided modes is very effective with this filamentary criss-cross geometry. Other peaks develop in the visible range of the spectrum. A difference, however is that theses peaks rapidly change their spectral location with the change of incidence plane orientation so that, in a disordered ensemble of incidences, their effect is distributed across the entire visible spectrum. However, the presence of these resonances in the UV and in the visible possibly leads to absorption in both ranges (UV and visible) of the spectrum. The absorption actually depends on the dissipation capabilities of the material traversed by the radiation, once entered in the filament. This issue is further discussed in the next section.

\section{Radiation absorption}

It is unclear whether the ultraviolet absorption can take place in the filament wall itself (which should then be characterized by a complex value of its refractive index) or whether the filament hollow core actually contains an absorbing agent. However, the absorption of these high-energy radiation (3-4 eV) is
likely to cause chemical damage and this mechanism should be confined to wavelengths beyond the high-energy end of the
visible spectrum.

A simple plausible mechanism of absorption involves the possible presence of water inside the hollow filaments. Pure water is known to absorb all electromagnetic radiations in the wavelengths range from kilometers to femtometers, except for those falling in the narrow window of the visible spectrum\cite{jackson-book-1999}. The absence of absorption in this biologically important range is just due to the lack of excitation mechanisms, between vibrational and electronic transitions of the water molecule. The absorption coefficient increases rapidly below 400 nm and will reach very high values (up to $10^{6}$ cm$^{-1}$) near 150 nm. In the near-UV range, molecular electronic excitation already contributes to destroy transparency so that water could well be the \bs agent'' suggested above. However, it is somewhat doubtful that the absorption coefficient of \textit{pure} water can already reach the values of $\alpha \simeq 10^{4}$ or $10^{5}$~cm$^{-1}$ needed to turn 10 $\mu$m thick filaments into efficient absorbers, above $300$ nm. For deeper UV radiation, UVB and UVC, there is no doubt that such a mechanism would be quite efficient but it remains questionable for UVA. On the other hand, it is also unlikely that the physiological liquid assumed to be filling the filament would actually be pure water. Unfortunately, the detection and analysis of the content of the filaments will not be an easy operation, due to the very small size of the pores and to the need for looking inside them \textit{in vivo}. An attempt would certainly be quite interesting and instructive, from the point of view of the present work. At present, an experimental argument against this mechanism is that the absorption is still clearly seen, without any weakening, in dry filaments, when the hair layer has been removed from the bract leaflets since several weeks.

\begin{figure}
\centerline{
\includegraphics[width=8cm]{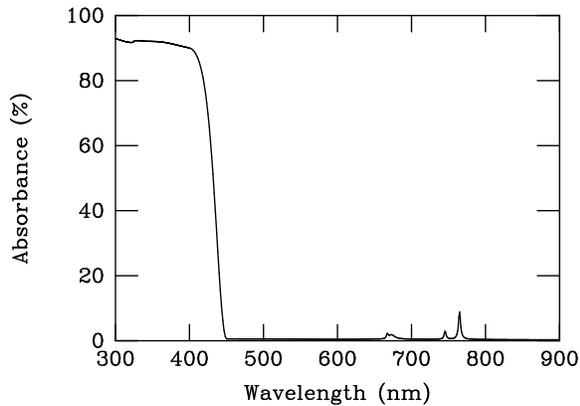}}
\caption{Absorption (in percents) of radiation on the model slab structured shown on Fig. \ref{fig11}, for an incidence transverse-electric wave impinging at 45$^{o}$ from normal in an incidence plane which makes a angle $\phi=30^{0}$ with the direction of the grating corrugation. The material in the model is given a complex dielectric constant, with an imaginary part which shifts from the value $\varepsilon'' = 0.4$ adequate for ultraviolet regions below 400 nm to $\varepsilon'' = 0.001$ adequate for the infrared and visible ranges, above 450 nm.} \label{fig13}
\end{figure}

Alternatively, the absorption can take place in the filament wall itself. In this case, the mechanism of energy injection into the guided modes would make the absorption by these tiny structures much more effective, by lengthening the absorbtion path of the light. However, on the basis of the model described by Fig. \ref{fig12}, the coupling to guided modes can occur in both the UV and in the visible parts of the spectrum, contradicting measurements~: the infrared and visible recordings have failed to show any significant absorption, while the UVA absorption approaches 100\%. As a consequence, a constant value of $\varepsilon''$ is not realistic and a dissipative response which rapidly shifts from high to low values at the UV onset would better represent the observed absorption. Without such a shift, the edelweiss bracts would absorb nearly as much in the visible range as it absorbs the near UV and, most likely, would appear dark. Fig. \ref{fig13} describes the computed absorption of the structure shown on Fig. \ref{fig11}, when the dielectric function imaginary part $\varepsilon''$ shifts from a value 0.4 in the UV range below 400 nm to a \bs transparent'' value of 0.001 maintained above 450 nm in the visible and infrared ranges. The molecular origin of this abrupt change is more than likely. As we can see, the physical origin of the whitening of the plant leaves involves the sharp disappearance of a dissipative response between the near-UV and visible ranges. This means that, besides the purely structural effects the \bs pigmentary'' (i.e. material-based rather than structural) mechanisms play a significant role in determining the bracts coloration. The detailed analysis of this aspect of the absorption process would require a detailed \textit{in vivo} study of the nature of the filament wall which is not yet available to date.

\section{Conclusion}
Photonic structures have been often found in nature\cite{parker-MT-2002,vukusic-Nat-2003}. Classic examples include the iridescence of insect cuticle\cite{berthier-book-2003,Parker-Nat-2003} (mainly butterflies, beetles, weevils, ...) or the changing colors of bird feathers\cite{Zi-PNAS-2003}. These photonic structures are most often the origin of bright signalling colors but, occasionally, they can be involved in the production of criptic colors, as in the butterfly \textit{Cyanophrys remus}\cite{biro-unp-2004}. In other cases, the photonic structure serves more physiological purposes, like the control of thermal energy exchange\cite{biro-PRE-2003}.

Photonic structures are not so common in plants, although some examples have been clearly described\cite{Bone-AO-1985,Lee-Nat-1991,Lee-AS-1997}. The observations presented here tend to show that \textit{Leontopodium nivale} could, from both the structural and optical points of view, be seen as another example. This plant has developed a layer of filamentary hair which covers the bracts surrounding the flowerheads and, with less density, the entire aerial part of the plant. This fleece protects from dehydration and cold, but also turns out to shield the covered living cells from harmful ultraviolet radiations. This protection is not obtained, as sometimes mentioned, by reflection but rather by absorption within the protective hair layer. The picture suggested here to explain the absorption involves a selective ultraviolet dissipation assisted by diffraction effects~: the fibrous structure of the external surface of the filaments not only provides a guiding support of ultraviolet radiation but also brings the roughness required for energy exchange between the incident waves and these guided modes.

With this structure, nature may have found here a clever solution to an engineering problem of wide application. Ultraviolet screening is of prime importance for the design of many functionalized materials. Examples can be given for packing materials, anti-ultraviolet sun screener for cosmetics, anti-ultraviolet powder for car or construction paints where the ultraviolet absorption is of paramount importance. Many of these applications rely on TiO$_{2}$ nanoparticles ($\sim 10-50$ nm) which may be difficult to handle, especially under the requirement of biocompatibility. The kind of structure developed by \textit{Leontopodium nivale} may indicate an interesting way to provide a strong ultraviolet absorption with larger (but structured) particles whose location can be much more easily stabilized.

\begin{acknowledgments}
This work was carried out with support from EU5 Centre of
Excellence ICAI-CT-2000-70029 and from the Inter-University
Attraction Pole (IUAP P5/1) on \bs Quantum-size effects in
nanostructured materials'' of the Belgian Office for Scientific,
Technical, and Cultural Affairs.

We acknowledge the use of Namur Interuniversity Scientific
Computing Facility (Namur-ISCF), a common project between the
Belgian National Fund for Scientific Research (FNRS), and the
Facult\'{e}s Universitaires Notre-Dame de la Paix (FUNDP).

M.R. has benefitted from a grant for research training at
the \textit{Laboratoire de Physique du Solide} (LPS) of the Facult\'{e}s
Universitaires Notre-Dame de la Paix (FUNDP).

ZV, LPB and KK wish to thank the Hungarian Academy of Sciences and
the Belgian FNRS for financial support.

V.L. was supported as research Fellow by the Belgian National Fund
for Scientific Research (FNRS). She is a recipient of a
postdoctoral fellowship from the Belgian-American Educational
Foundation.

The authors thank Amand Lucas (FUNDP) and Paul Thiry (FUNDP) for critically
reading the manuscript and for stimulating discussions.
\end{acknowledgments}

\end{document}